\documentclass[aps,twocolumn,eqsecnum,showpacs,amsmath,footinbib]{revtex4-1}
\usepackage{color,graphicx}
\usepackage{graphicx}
\usepackage{amsfonts,amssymb,theorem}
\usepackage{ulem}
\newcommand{\D}{{\rm d}}

{\theorembodyfont{\upshape}
}
{\theorembodyfont{\upshape}
}
{\theorembodyfont{\upshape}
}
{\theorembodyfont{\upshape}
}
{\theorembodyfont{\upshape}
}
{\theorembodyfont{\upshape}
}

\newcommand{\dalm}{\kern1pt\vbox{\hrule height 0.9pt\hbox{\vrule width
0.9pt\hskip 2.5pt\vbox{\vskip 5.5pt}\hskip 3pt\vrule width 0.3pt}\hrule height
0.3pt}\kern1pt}

\begin{document}

\title{
Thakurta metric does not describe a cosmological black hole
}

\author{Tomohiro Harada${}^a$}
\email{harada@rikkyo.ac.jp}
\author{Hideki Maeda${}^b$}
\email{h-maeda@hgu.jp}
\author{Takuma Sato${}^a$}
\email{stakuma@rikkyo.ac.jp}


\affiliation{
${}^a$ Department of Physics, Rikkyo University, Tokyo 171-8501, Japan}
\affiliation{ ${}^b$ Department of Electronics and Information Engineering, Hokkai-Gakuen University, Sapporo 062-8605, Japan}

\date{\today}

\begin{abstract}
Recently, the Thakurta metric has been adopted as a model of primordial black holes. We show that the spacetime described by this metric 
has neither black-hole event horizon nor black-hole 
trapping horizon and involves the violation of 
all the standard energy conditions as a solution of the Einstein equation. 
Therefore, this metric does not describe 
a cosmological black hole in the early universe.
It is pointed out that a contradictory claim by the other group stems 
from an incorrect choice of sign.
\end{abstract}



\maketitle


\section{Introduction}
Primordial black holes (PBHs)~\cite{Carr:1974nx,Carr:1975qj} have attracted intense attention as 
it has been revealed that 
a considerable fraction of binary black holes discovered by gravitational wave
observations by LIGO and Virgo have masses in the
range of several tens of solar masses. In fact, it is proposed that
such very massive black holes might be of cosmological origin~\cite{Sasaki:2016jop,Clesse:2016vqa,Bird:2016dcv}. 
Recently, in the data observed by LIGO and Virgo, 
Franciolini et al.~\cite{Franciolini:2021tla} searched for a subpopulation of PBHs and found that the statistical evidence significantly depends on the assumed astrophysical models.

To discuss the physical properties of PBHs, 
Boehm et al.~\cite{Boehm:2020jwd} and 
Picker~\cite{Picker:2021jxl} adopt
the Thakurta metric~\cite{Thakurta1983,Faraoni:2018xwo}.
These works are criticized by H\"{u}tsi et al.~\cite{Hutsi:2021nvs} as
this spacetime entails accretion history which is highly unlikely from a physical point of view.
This controversy continues by a further comment~\cite{Boehm:2021kzq} and a reply~\cite{Hutsi:2021vha}.
In a recent article~\cite{Kobakhidze:2021rsh}, the authors have claimed that the Thakurta spacetime admits a future outer trapping horizon in the Painlev\'{e}-Gullstrand-like coordinates, so that it may be interpreted as a cosmological black hole.

In this letter, we terminate the controversy on the Thakurta metric to turn the discussion to the right direction to seek for a mathematical model of PBHs.
Although some of the main results in this letter are included in~\cite{Mello:2016irl}, where the causal structures of the Thakurta metric are 
properly analyzed in more general contexts, we will also study the energy conditions of the corresponding matter field and the properties of a trapping horizon in more detail.
Furthermore, we will point out an error in the contradictory argument in Ref.~\cite{Kobakhidze:2021rsh}.
Note that the Thakurta metric is different from
McVittie's solutions~\cite{Faraoni:2018xwo,McVittie:1933zz}, whose causal structures are 
analyzed in detail in~\cite{Nolan:1998xs,Nolan:1999kk,Nolan:1999wf,Kaloper:2010ec}.
Our conventions for curvature tensors are $[\nabla _\rho ,\nabla_\sigma]V^\mu ={{\cal R}^\mu }_{\nu\rho\sigma}V^\nu$ and ${\cal R}_{\mu \nu }={{\cal R}^\rho }_{\mu \rho \nu }$.
The signature of the Minkowski spacetime is $(-,+,+,+)$, and Greek indices run over all spacetime indices.
We use the units in which $c=G=1$.

\section{Preliminaries}
Consider the most general four-dimensional spherically symmetric spacetime $(M^4,g_{\mu\nu})$ given by
\begin{align}
\D s^2 =g_{AB}(y)\D y^A\D y^B +R^2(y)\D\Omega^{2},
\label{eq:ansatz}
\end{align}
where $y^A~(A=0,1)$ are coordinates in a two-dimensional Lorentzian spacetime $(M^2, g_{AB})$ and $\D\Omega^{2}:=\D\theta^2+\sin^2\theta\D\phi^2$.
The Misner-Sharp quasi-local mass $m_{\rm MS}$~\cite{Misner:1964je} for the metric (\ref{eq:ansatz}) is defined by
\begin{align}
m_{\rm MS} := \frac{1}{2}R\left\{1-(DR)^2\right\},\label{MS-mass}
\end{align}
where $(DR)^2:=g^{AB}(D_A R)(D_B R)$ and $D_A$ is the covariant derivative on $(M^2, g_{AB})$.
Note that $m_{\rm MS}$ and the areal radius $R(y)(\ge 0)$ are scalars on $(M^2, g_{AB})$.

In this letter, we adopt spherical slicings to identify trapped regions and trapping horizons defined by Hayward~\cite{Hayward:1993wb}.
Let $\boldsymbol{k}$ and $\boldsymbol{l}$ be
two independent future-directed radial null vectors such that 
$k^\mu(\partial/\partial x^\mu)=k^A(\partial/\partial y^A)$ and $l^\mu(\partial/\partial x^\mu)=l^A(\partial/\partial y^A)$ with
\begin{align}
k_\mu k^\mu=l_\mu l^\mu=0,\qquad k_\mu l^\mu=-1. \label{kl2}
\end{align}
The expansions along those null vectors are given by
\begin{align}
\theta_+:=&{\cal A}^{-1}{\cal L}_+{\cal A}=2R^{-1}k^AD_AR,\label{exp+}\\
\theta_-:=&{\cal A}^{-1}{\cal L}_-{\cal A}=2R^{-1}l^AD_AR, \label{exp-}
\end{align}
where ${\cal L}_+:=k^AD_A$ and ${\cal L}_-:=l^AD_A$.
Here ${\cal A}:=4\pi R^2$ is the surface area with the areal radius $R$.
By the expression $\theta_+\theta_-=-2R^{-2}(DR)^2$ and Eq.~(\ref{MS-mass}), an {\it untrapped (trapped) region} defined by $\theta_{+}\theta_{-}<(>)0$ is given by $(D R)^2>(<)0$, or equivalently $R>(<)2m_{\rm MS}$.
A {\it marginal surface} defined by $\theta_{+}\theta_{-}=0$ is given by $(D R)^2=0$, or equivalently $R=2m_{\rm MS}$.

A {\it trapping horizon} is the closure of a hypersurface foliated by marginal surfaces~\cite{Hayward:1993wb}.
Using the degrees of freedom in interchanging $\theta_+$ and $\theta_-$, one may set $\theta_{+}=0$ on a trapping horizon without loss of generality.
Then, a trapping horizon is classified according to the signs of $\theta_-$ and ${\cal L}_-\theta_{+}$ there as summarized in Table~\ref{table:TH-class}.
\begin{table}[htb]
\begin{center}
\caption{\label{table:TH-class} Types of trapping horizons given by $\theta_+=0$.}
\scalebox{1.00}{
\begin{tabular}{|c|c|c|c|c|}
\hline
& $+$ & $0$ & $-$ \\ \hline\hline
~$\theta_-$~ & ~Past~ & ~Bifurcating~ & ~Future~ \\ \hline
~${\cal L}_-\theta_{+}$~ & ~Inner~ & ~Degenerate~ & ~Outer~ \\
\hline
\end{tabular}
}
\end{center}
\end{table}

Among all the types of trapping horizons, it is a {\it future outer trapping horizon} that is associated with a black hole~\cite{Hayward:1993wb}.
This is based on (i) $\theta_-|_\Sigma<0$ meaning that the ingoing null rays converge on the trapping horizon $\Sigma$ and (ii) ${\cal L}_-\theta_{+}|_\Sigma<0$ meaning that the outgoing null rays are instantaneously parallel on $\Sigma$ but diverging just outside $\Sigma$ and converging just inside~\cite{Hayward:1993wb,Hayward:1997jp}.
In contrast, a past inner trapping horizon and a past outer trapping horizon correspond to a cosmological horizon and a white-hole horizon, respectively.

\section{Global structure of the Thakurta spacetime}

The Thakurta metric is given by
\begin{align}
\label{eq:Thakurta}
\begin{aligned}
&\D s^2=a^{2}(\eta)\left[-f(r)\D \eta^2+\frac{1}{f(r)}\D r^2+r^2\D \Omega^2\right], \\
&f(r):=1-\frac{2M}{r},
\end{aligned}
\end{align}
where $M$ is a constant~\cite{Thakurta1983,Faraoni:2018xwo}. We assume $M>0$ throughout this letter.
The spacetime approaches the flat Friedmann-Lemaitre-Robertson-Walker (FLRW) 
solution with a conformal time $\eta$ as $r\to \infty$.
If the scale factor is given by $a(\eta)=a_{0}\eta^{\alpha}$, where $a_{0}$ and $\alpha$ are positive constants, 
the asymptotic metric is that of the flat FLRW solution with a perfect fluid 
that obeys the equation of state $p=w\rho$ $(w>-1/3)$ 
through the relation $\alpha=2/(1+3w)$, corresponding to a decelerated cosmological expansion in the domain $\eta>0$.

The Kretschmann invariant $K:={\cal R}_{\mu\nu\rho\sigma}{\cal R}^{\mu\nu\rho\sigma}$ for the metric (\ref{eq:Thakurta}) is calculated to give
\begin{equation}
K=\frac{4[3\alpha^2(1+\alpha^2)r^{8}+4M^2\eta^2\{3\eta^{2}(r-2M)^2 -\alpha^2 r^4\}]}{a_0^4r^6(r-2M)^2\eta^{4(1+\alpha)}}, \label{K-div1}
\end{equation} 
which shows that there is a scalar polynomial (s.p.) curvature singularity at each of $r=0$ and $\eta=0$. $\eta=0$ with $r>2M$ is a spacelike singularity.
However, the property of $r=2M$ is subtle.

The line element (\ref{eq:Thakurta}) can be written as
\begin{equation}
\D s^2=a^{2}(\eta)\left[f(r)(-\D \eta^2+\D {r^*}^2)+r^{2}\D\Omega^{2}\right], 
\end{equation}
where $r^{*}:=r+2M\ln[(r-2M)/{2M}]$.
Since $r\to 2M$ corresponds to $r^*\to -\infty$, it is a null boundary in the Penrose diagram.

Now we show that future-directed ingoing radial null geodesics are incomplete at the s.p. curvature singularity at $(r,\eta)=(2M,\infty)$.
In the Thakurta metric (\ref{eq:Thakurta}), an ingoing radial null geodesic $\gamma$ satisfies $u_{+}=u_{+(0)}=\mbox{const}$, where 
$u_{\pm}:=\eta\pm r^{*}$.
The Kretschmann invariant in the limit of $r\to 2M$ along $\gamma$
is then calculated to
\begin{equation}
\lim_{r\to 2M}K|_{\gamma}= \lim_{r^{*}\to -\infty}\biggl(-\frac{\alpha^{2}e^{-(r^{*}-2M)/M}}
{M^{4}a_{0}^{4}(-r^{*})^{2+4\alpha}}\biggl)=-\infty.
\end{equation}
Let $\boldsymbol{p}$ be a tangent vector of $\gamma$. Since the spacetime admits a conformal Killing vector $\xi^\mu=(1,0,0,0)$, there exists a conserved quantity $C=-\xi^{\mu}p_{\mu}=a^{2}(\eta)f(r)\dot{\eta}$ along $\gamma$, where the dot denotes the derivative with respect to the affine parameter $\lambda$. We can make $C=1$ by an affine transformation and then we find $\D\lambda=-a^{2}(\eta)\D r$ through the relation $p^{\mu}p_{\mu}=0$.
The value of $\lambda$ with which $\gamma$
terminates at $r=2M$ is then given by 
\begin{equation}
\lambda-\lambda_{0}=\int_{-\infty}^{r_{0}^{*}}
a^{2}(u_{+(0)}-r^{*})f(r)\D r^{*}.
\label{eq:affine}
\end{equation}
The interval of the integral on the right-hand side of Eq.~(\ref{eq:affine}) can be divided to $(-\infty,-L]$ and $[-L,r_{0}^{*}]$.
For any $\epsilon \in (0,1/2)$, there exists 
sufficiently large but finite $L$ such that
\begin{equation}
(u_{+(0)}-r^{*})^{2\alpha}f(r)<(2M)^{2\alpha}e^{(1-\epsilon)(r^{*}-2M)/(2M)}
\end{equation}
for all $ r^{*}<-L$. Therefore, if $a(\eta)=a_{0}\eta^{\alpha}$, the integral over the first interval converges to a finite limit value, while that over the second is trivially finite.
This means that the ingoing radial null geodesics are future incomplete at 
the s.p. curvature singularity at $(r,\eta)=(2M,\infty)$. 

\begin{figure}[htbp]
\includegraphics[width=0.3\textwidth]{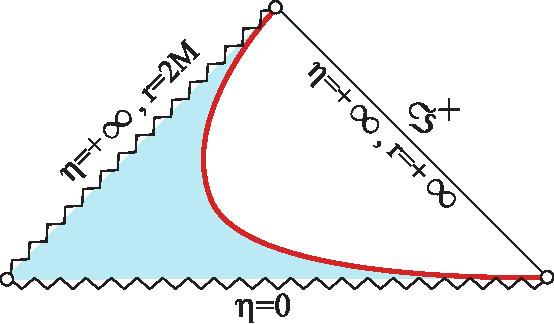}
\includegraphics[width=0.3\textwidth]{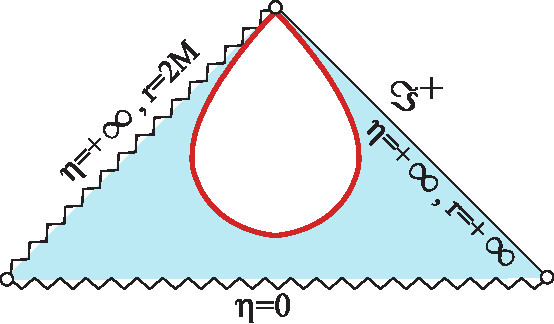}
\caption{
Penrose diagrams of the Thakurta spacetime with $a(\eta)\propto \eta^{\alpha}$ for $0<\alpha\le 1$ (top) and $\alpha> 1$ (bottom). 
Solid curves denote trapping horizons, which are past trapping horizons, 
and the shaded regions are past trapped. 
There is neither event horizon nor future outer trapping horizon in the spacetime. 
}
\label{fg:Thakurta_Penrose}
\end{figure}

As a result, the Penrose diagrams of the Thakurta spacetime 
with $a(\eta)\propto \eta^{\alpha}$ $(\alpha>0)$ are given by Fig.~\ref{fg:Thakurta_Penrose}. (See~\cite{Harada:2018ikn,Harada:2021yul} 
for those of the corresponding FLRW spacetimes.)
This is a maximal extension of the spacetime.
Clearly, there is no black-hole event horizon.
Nevertheless, the spacetime might be interpreted as a black hole by an alternative definition in terms of a trapping horizon~\cite{Hayward:1993wb,Hayward:1994bu} or an apparent horizon~\cite{he1973,wald1983}. 
In fact, the Thakurta spacetime possesses a trapping horizon in the spherical foliation of the spacetime. 
However, this trapping horizon is not of black-hole type as we will show below.

\section{Trapping horizon}

With $R=ar$, the Misner-Sharp mass (\ref{MS-mass}) for the Thakurta spacetime is computed to give
\begin{equation}
m_{\rm MS}=a\left(M+\frac{{\cal H}^{2}r^{3}}{2f(r)}\right),\label{def-MS}
\end{equation}
where ${\cal H}:=a'/a$.
A prime denotes the derivative with the argument $\eta$.
The location of a trapping horizon is determined by $R=2m_{\rm MS}$, or 
\begin{equation}
f(r)={\cal H}r, \label{eq:TH}
\end{equation}
where $r>2M$ is assumed.
With ${\cal H}=\alpha/\eta$, Eq.~(\ref{eq:TH}) is solved for $\eta$ to give
\begin{equation}
\eta=\frac{\alpha r^2}{r-2M}=:\eta_{\rm TH}(r).\label{eta-TH}
\end{equation}
The function $\eta=\eta_{\rm TH}(r)$ has a single local minimum $\eta=8\alpha M$ at $r=4M$.

In fact, the Thakurta metric (\ref{eq:Thakurta}) 
with $a(\eta)\propto \eta^{\alpha}$ $(\alpha>0)$ admits only a {\it past} trapping horizon
in the domain $\eta>0$, which is associated with a white-hole or a cosmological horizon.
To see this, we adopt the following choice for 
$\boldsymbol{k}$ and $\boldsymbol{l}$:
\begin{align}
k^\mu\frac{\partial}{\partial x^\mu}=&k^A\frac{\partial}{\partial y^A}=\frac{1}{\sqrt{2}a}\biggl(\frac{1}{\sqrt{f}}\frac{\partial}{\partial \eta}+\sqrt{f}\frac{\partial}{\partial r}\biggl),\\
l^\mu\frac{\partial}{\partial x^\mu}=&l^A\frac{\partial}{\partial y^A}=\frac{1}{\sqrt{2}a}\biggl(\frac{1}{\sqrt{f}}\frac{\partial}{\partial \eta}-\sqrt{f}\frac{\partial}{\partial r}\biggl),
\end{align}
which satisfy Eq.~(\ref{kl2}).
By Eqs.~(\ref{exp+}) and (\ref{exp-}), the 
null expansions along $\boldsymbol{k}$ and $\boldsymbol{l}$
are given by $\theta_{+}$ and $\theta_{-}$, respectively, where
\begin{equation}
\theta_{\pm}=\frac{\sqrt{2}\left[{\cal H}r\pm f(r)\right]}{ar\sqrt{f}}.\label{eq:theta-pm2}
\end{equation}
Because $\theta_+>\theta_-$ holds, $\boldsymbol{k}$ and $\boldsymbol{l}$
are outgoing and ingoing, respectively.
Equation~(\ref{eq:theta-pm2}) shows that $\theta_{+}>0$ and $\theta_{-}=0$ hold on the trapping horizon~(\ref{eq:TH}).
For the further classification of a trapping horizon, we need to use
\begin{equation}
{\cal L}_+\theta_{-}|_{\theta_-=0}=-\frac{(1-\alpha)(r-2M)+2\alpha M}{\alpha a^2r^3}.\label{eq:theta-pm2-d}
\end{equation}

Since $\theta_{+}|_{\theta_{-}=0}>0$ holds, our trapping horizon (\ref{eta-TH}) is a {\it past} trapping horizon. 
Then, by Eq.~(\ref{eq:theta-pm2-d}), it is an {\it outer} trapping horizon for $0<\alpha \le 1$.
For $\alpha>1$, it is an outer, degenerate, and inner trapping horizons in the regions of $r-2M<\beta_2$, $r-2M=\beta_2$, and $r-2M>\beta_2$, respectively, 
where 
\begin{equation}
\beta_2:=\frac{2\alpha M}{\alpha-1}. \label{Delta-r}
\end{equation}
In the region of $0<\eta<\eta_{\rm TH}(r)$, both $\theta_{+}$ and $\theta_{-}$ are positive, meaning that the region is {\it past} trapped. 
These results are summarized in Table~\ref{table:TH-type}.
\begin{table}[htb]
\begin{center}
\caption{\label{table:TH-type} Types of the trapping horizon in the Thakurta spacetime with $a(\eta)\propto \eta^{\alpha}$ $(\alpha>0)$ for the different ranges of $r-2M$.}
\begin{tabular}{|c|c|c|c|c|}
\hline
& Past outer & Past degenerate & Past inner \\ \hline\hline
$0<\alpha\le 1$ & $(0,\infty)$ & $\emptyset$ & $\emptyset$ \\ \hline
$\alpha>1$ & $(0,\beta_2)$ & $\beta_2$ & $(\beta_{2},\infty)$ \\ 
\hline
\end{tabular} 
\end{center}
\end{table} 

We also check the signature of the trapping horizon as another important property. 
The line element on the trapping horizon (\ref{eta-TH}) is given by 
\begin{align}
\label{eq:signature}
\begin{aligned}
\D s^{2}_{\rm TH}=&a^{2}(\eta_{\rm TH}(r))\left(\frac{\Pi(r)}{r^{2}f^{3}}\D r^{2}+r^{2}\D\Omega^{2}\right),\\
\Pi(r):=&[(1-\alpha)(r-2M)+2\alpha M]\\
&\times[(1+\alpha)(r-2M)-2\alpha M],
\end{aligned}
\end{align}
which clarifies the locations of timelike, null, and spacelike portions of the trapping horizon.
The results are summarized in Table~\ref{table:TH-sig}.
We can see that 
the signature changes at $r-2M=\beta_{1}$ and $\beta_{2}$, where 
\begin{align}
\beta_1:=\frac{2\alpha M}{\alpha+1}.\label{def-beta1}
\end{align}
We notice that for $\alpha=1$, $u_{-}\to -\infty$ and $u_{+}\to+\infty$ 
holds as $r\to \infty$ along the trapping horizon $\eta=\eta_{\rm TH}(r)$.
This limit corresponds to spacelike infinity $i^0$ in the Penrose diagram.
\begin{table}[htb]
\begin{center}
\caption{\label{table:TH-sig} Signatures of the trapping horizon for the different ranges of $r-2M$.}
\begin{tabular}{|c|c|c|c|}
\hline
& Timelike & $\quad$ Null $\quad$ & ~~Spacelike~~ \\ \hline\hline
$0<\alpha\le 1$ & $(0,\beta_{1})$ & $\beta_{1}$ & $(\beta_{1},\infty)$ \\ \hline
$1<\alpha$ & $(0,\beta_{1}), (\beta_{2},\infty)$ & $\beta_{1}, \beta_{2}$ & $(\beta_{1},\beta_{2})$ \\ 
\hline
\end{tabular} 
\end{center}
\end{table} 

\begin{figure}[htbp]
\includegraphics[width=0.4\textwidth]{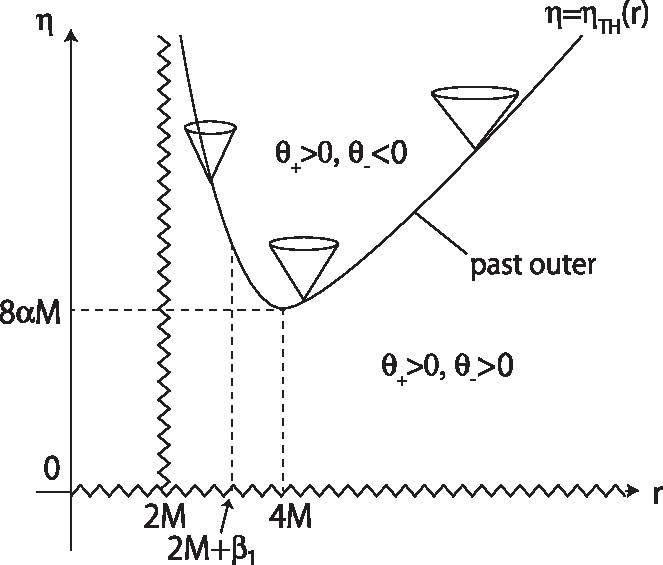}
\includegraphics[width=0.4\textwidth]{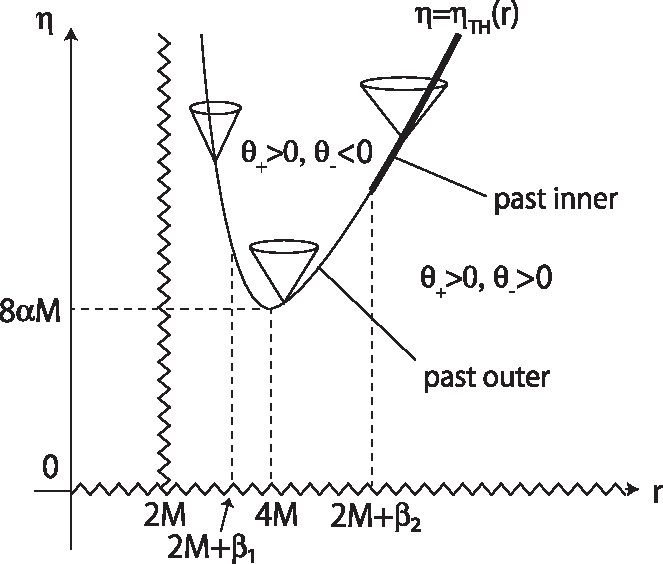}
\caption{
$r\eta$ planes for the Thakurta spacetime with $a(\eta)\propto \eta^{\alpha}$ for (top) $0<\alpha\le 1$ and 
(bottom) $\alpha>1$. 
Several future light cones are put to clarify the signature of the trapping horizon $\eta=\eta_{\rm TH}(r)$.
The regions of $0<\eta<\eta_{\rm TH}(r)$ with $r>2M$ are past trapped.
}
\label{fg:TH-r-eta1}
\end{figure}

We present all the results obtained so far in the
$r\eta$ plane in Fig.~\ref{fg:TH-r-eta1}.
The orbits of the trapping horizons are also presented in Fig.~\ref{fg:Thakurta_Penrose}.
We find that for all $\alpha>0$, there exists a past outer trapping horizon 
located at $0<r-2M<\beta_{1}$, which is a timelike hypersurface.
It is known that an outer (inner) trapping horizon is nontimelike (nonspacelike) under the null energy condition (NEC) in general relativity~\cite{Hayward:1993wb,Nozawa:2007vq}.
Therefore, if we adopt the Thakurta metric (\ref{eq:Thakurta}) as a solution of the Einstein equation, the corresponding matter field violates the NEC on 
the trapping horizon located at $0<r-2M<\beta_{1}$ for any $\alpha>0$.

In fact, the corresponding energy-momentum tensor $T_{\mu\nu}(:=G_{\mu\nu}/(8\pi))$ is of the Hawking-Ellis type IV in the region of $\eta>(\alpha+1)r^2/(2M)$, which violates all the standard energy conditions~\cite{he1973,Maeda:2018hqu,Martin-Moruno:2017exc}.
To prove this, we compute $G^{(a)(b)}:=G^{\mu\nu}E^{(a)}_{\mu}E^{(b)}_{\nu}$ with the following orthonormal basis one-forms:
\begin{align}
\begin{aligned}
&E_\mu^{(0)}\D x^\mu=-a(\eta)f(r)^{1/2}\D \eta,\\
&E_\mu^{(1)}\D x^\mu=a(\eta)f(r)^{-1/2}\D r, \\
&{E}_\mu^{(2)}\D x^\mu=a(\eta)r\D \theta,\\
&{E}_\mu^{(3)}\D x^\mu=a(\eta)r\sin\theta\D \phi.
\end{aligned}
\end{align}
All the non-zero components $G^{(a)(b)}$ are
\begin{align}
\begin{aligned}
&G^{(0)(0)}=\frac{3}{f}\frac{{a'}^2}{a^4},\qquad G^{(0)(1)}=G^{(1)(0)}=\frac{2M}{r^2f}\frac{a'}{a^3},\\
&G^{(1)(1)}=G^{(2)(2)}=G^{(3)(3)}=\frac{1}{f}\biggl(-2\frac{a''}{a^3}+\frac{{a'}^2}{a^4}\biggl),
\end{aligned}
\end{align}
which give
\begin{align}
{\cal D}:=&(G^{(0)(0)}+G^{(1)(1)})^2-4(G^{(0)(1)})^2 \nonumber\\
=&\frac{4\alpha^2}{\eta^2f^2a^4}\biggl(\frac{\alpha+1}{\eta}+\frac{2M}{r^2}\biggl)\biggl(\frac{\alpha+1}{\eta}-\frac{2M}{r^2}\biggl).\label{D-Thakurta2}
\end{align}
Then, by Lemma~1 in ~\cite{Maeda:2022vld}, the corresponding energy-momentum tensor $T_{\mu\nu}$ is of the Hawking-Ellis type I, II, and IV in the regions of ${\cal D}>0$, ${\cal D}=0$, and ${\cal D}<0$, respectively.
Thus, by Eq.~(\ref{D-Thakurta2}), $T_{\mu\nu}$ is of type IV in the region of $\eta>(\alpha+1)r^2/(2M)$ and therefore violates all the standard energy conditions there.
On the trapping horizon (\ref{eta-TH}), $T_{\mu\nu}$ is of type IV in the region of $0<r-2M<\beta_1$, where $\beta_1$ is defined by Eq.~(\ref{def-beta1}).
This is the reason of the unusual behavior of the trapping horizon in this region.

\section{Trapping horizon in Painlev\'{e}-Gullstrand-like coordinates}

In Ref.~\cite{Kobakhidze:2021rsh}, the authors claim that the trapping horizon in the Thakurta spacetime considered in this letter can be a future outer trapping horizon in Painlev\'{e}-Gullstrand-like coordinates on $(M^2,g_{AB})$ and then the spacetime may be interpreted as a cosmological black hole.
However, this claim cannot be true. In fact,  
the location of a trapping horizon and its type are invariant 
as long as a spherical foliation is adopted
because $\theta_{\pm}$ and ${\cal L}_{\pm}\theta_{\mp}$ 
are scalars on $(M^2,g_{AB})$~\cite{Faraoni:2016xgy}.
Moreover, the signs of $\theta_{\pm}$ and ${\cal L}_{\pm}\theta_{\mp}$ 
are independent from the choice of $\boldsymbol{k}$ and $\boldsymbol{l}$. 
(See Proposition 1 in Ref.~\cite{Sato:2022yto}.)
In contrast, it has been established that {\it non-spherical} foliations may miss  
trapped surfaces~\cite{Wald:1991zz,Schnetter:2005ea}.

Here we find out an error in the argument in Ref.~\cite{Kobakhidze:2021rsh}.
First we outline the argument made in Ref.~\cite{Kobakhidze:2021rsh}.
The authors start from the Thakurta metric in the following coordinates:
\begin{align}
&\D s^2=-f(r)\D t^2+{\bar a}^2(t)\left(\frac{1}{f(r)}\D r^2+r^2\D \Omega^2\right),\label{Thakurta-origin}
\end{align}
where $f(r)$ is defined by Eq.~(\ref{eq:Thakurta}). 
They introduce the following scalar function $F$ on $(M^2,g_{AB})$ defined by the areal radius $R$ and the Misner-Sharp mass (\ref{MS-mass}):
\begin{align}
F:=&1-\frac{2m_{\rm MS}}{R}. \label{def-F}
\end{align}
A trapping horizon is given by $F=0$.

By a coordinate transformation from $(t,r)$ to $(t,R)$ with 
$R={\bar a}(t)r$, one obtains
\begin{align}
\label{T2}
\begin{aligned}
\D s^2=&-{\bar f}\D t^2+\frac{1}{{\bar f}}\left(\D R-HR\D t\right)^2+R^2\D \Omega^2,\\
{\bar f}:=&1-\frac{2M{\bar a}(t)}{R},\qquad H:=\frac{{\bar a}_{,t}}{{\bar a}},
\end{aligned}
\end{align}
where a comma denotes a partial differentiation
and $H>0$ corresponds to the case where the asymptotic region is the expanding FLRW universe.
The function $F$ is computed to give
\begin{align}
F={\bar f}-\frac{H^2R^2}{{\bar f}}.\label{F1}
\end{align}
Then, we perform another coordinate transformation from $(t,R)$ to $(\tau,R)$, 
where $\tau$ is defined such that $t=t(\tau,R)$ satisfies
\begin{align}
{\bar f}=-{\bar f}^2{t_{,R}}^2+(1-HRt_{,R})^2\label{pde1}
\end{align}
to give $g_{RR}=1$.
Note that the coordinates $(\tau,R)$ are 
the Painlev\'{e}-Gullstrand-like coordinates.
The resulting metric is
\begin{align}
\D s^2=&-{t_{,\tau}}^2F\D \tau^2-2t_{,\tau}\left(Ft_{,R}+
\frac{HR}{{\bar f}}\right)\D\tau\D R \nonumber \\
&+\D R^2+R^2\D \Omega^2,\label{T3}
\end{align}
where we used Eq.~(\ref{F1}).
Equations~(\ref{F1}) and (\ref{pde1}) show
\begin{align}
\left(Ft_{,R}+\frac{HR}{{\bar f}}\right)^2=1-F,\label{key1}
\end{align}
which gives 
\begin{align}
Ft_{,R}+
\frac{HR}{{\bar f}}=\varepsilon\sqrt{1-F},\label{key2}
\end{align}
where $\varepsilon=\pm 1$.
Hence, the metric (\ref{T3}) reduces to
\begin{align}
\D s^2=&-t_{,\tau}^2F\D \tau^2 -2\varepsilon t_{,\tau}\sqrt{1-F}\D\tau\D R
\nonumber \\
&~~~~~+\D R^2+R^2\D \Omega^2.
\label{T4}
\end{align}

Now we are prepared to see that the contradictory argument 
in Ref.~\cite{Kobakhidze:2021rsh} stems from an incorrect choice of $\varepsilon$.
The authors chose $\varepsilon=-1$ in the metric (\ref{T4}) 
as seen in Eq. (21) in Ref.~\cite{Kobakhidze:2021rsh} .
(Note that the metric signature $(+,-,-,-)$ was adopted 
in Ref.~\cite{Kobakhidze:2021rsh}.)
However, 
Eq.~(\ref{key2}) with $\varepsilon=-1$ and $H>0$ leads to a contradiction on a trapping horizon $F=0$ in the region of ${\bar f}>0$, where $t_{,R}$ is assumed to be finite. Clearly, one has to choose $\varepsilon=1$ instead
\footnote{The sign of $\varepsilon$ is invariant under the transformation from $\tau$ to $-\tau$.}.

Let us confirm that the trapping horizon in the coordinate system (\ref{T4}) with $\varepsilon=1$ is of the past type.
We adopt the following two independent future-directed radial null vectors:
\begin{align}
k^\mu\frac{\partial}{\partial x^\mu}=&\frac{1}{\sqrt{2}}\biggl(\frac{1}{{t_{,\tau}}}\frac{\partial}{\partial {\tau}}+(1+\varepsilon \sqrt{1-F})\frac{\partial}{\partial R}\biggl),\\
l^\mu\frac{\partial}{\partial x^\mu}=&\frac{1}{\sqrt{2}}\biggl(\frac{1}{{t_{,\tau}}}\frac{\partial}{\partial {\tau}}-(1-\varepsilon \sqrt{1-F})\frac{\partial}{\partial R}\biggl),
\end{align}
which satisfy Eq.~(\ref{kl2}).
From Eqs.~(\ref{exp+}) and (\ref{exp-}), 
the null expansions along $\boldsymbol{k}$ and 
$\boldsymbol{l}$ are given by
\begin{align}
\theta_{+}=\frac{\sqrt{2}(1+\varepsilon \sqrt{1-F})}{R},\quad \theta_{-}=-\frac{\sqrt{2}(1-\varepsilon \sqrt{1-F})}{R},
\end{align}
respectively.
Since $\theta_+>\theta_-$ holds independently of the value of $\varepsilon$, $\boldsymbol{k}$ and $\boldsymbol{l}$ 
are outgoing and ingoing, respectively.
On a trapping horizon $F=0$ with $\varepsilon=1$, we have $\theta_+>0$ and $\theta_-=0$, so that it is of the past type.
In contrast, with $\varepsilon=-1$, one obtains $\theta_+=0$ and $\theta_-<0$ on the trapping horizon and then it is of the future type.
However, by Eq.~(\ref{key2}), $\varepsilon=-1$ is allowed only with $H<0$. 
Namely, a future trapping horizon may be possible only 
in the case where the asymptotic region is the collapsing FLRW universe, which is of no interest in the present context.

\section{Concluding remarks}

We emphasize that the Thakurta metric does not describe a cosmological black hole in spite of subtlety in defining cosmological black holes.
Let us consider an observer O located outside a black-hole event horizon 
at finite distance and time, not at future null infinity nor timelike infinity.
This is an observer usually presumed in observational cosmology.
By definition, such an observer never knows the 
existence of an event horizon,  
even though the causal nature of the $r=2M$ surface, such as whether it is finitely far or not 
and regular or not, crucially depends on the asymptotic behavior of $a(\eta)$ in
the infinite future $\eta\to \infty$ (cf. \cite{Nakao:2018knn}).

What is even worse, if the universe is not asymptotically flat FLRW, it may not 
be possible to define future null infinity nor an event horizon.
These facts suggest that, at least from an observational point of view, we should give up the definition of a cosmological black hole in terms of an event horizon. 
Instead, for example, we may practically define a cosmological black hole by the existence of an ``almost'' future outer trapping horizon, a hypersurface foliated by ``almost'' marginal surfaces on which $\theta_{-}<0$ and
$\theta_{+}$ is infinitesimally small positive, and ${\cal L}_{-}\theta_{+}<0$ are satisfied, 
that can be observed in principle by O.
Even in such a rather weaker definition, the Thakurta spacetime 
cannot be interpreted as a cosmological black hole because both $\theta_{+}$ 
and $\theta_{-}$ are always positive in the vicinity of $r=2M$ surface.

To conclude, the Thakurta metric (\ref{eq:Thakurta}) with $a(\eta)\propto \eta^{\alpha}$ $(\alpha>0)$ does not describe a cosmological black hole 
in the early universe, with a possible exceptional case, where the cosmological expansion is accelerated.

\acknowledgments

T. H. is grateful to N. Tanahashi and J.~Yang 
for fruitful discussion.
H. M. is grateful to A. Kobakhidze for email communications.
This work was partially supported by MEXT Grants-in-Aid for Scientific Research/JSPSKAKENHI Grants Nos. JP19K03876, JP19H01895, and JP20H05853 (T.H.).


\end{document}